\documentclass[12pt,a4paper,fleqn]{article}

\usepackage{graphicx}
\usepackage{amsmath}
\usepackage{amssymb}
\usepackage{longtable}

\newcommand{\be}{\begin{equation}}
\newcommand{\ee}{\end{equation}}
\newcommand{\bea}{\begin{eqnarray}}
\newcommand{\eea}{\end{eqnarray}}
\newcommand{\bi}{\begin{itemize}}
\newcommand{\ei}{\end{itemize}}
\newcommand{\ben}{\begin{enumerate}}
\newcommand{\een}{\end{enumerate}}
\newcommand{\bt}{\begin{tabbing}}
\newcommand{\et}{\end{tabbing}}

\newcommand{\calO}{{\mathcal O}}
\newcommand{\cA}{c_{\rm A}}
\newcommand{\ca}{c_{\rm A}}
\newcommand{\nf}{N_{\rm f}}
\newcommand{\Tr}{{\rm Tr}}

\title{ 
\begin{flushright}
\small CERN-PH-TH/2007-030 \\
DESY 07-005\\
KEK-CP-192\\
UTHEP-539\\
COLO-HEP-522\\[15mm]
\end{flushright}
Non-perturbative improvement of the axial current 
with three dynamical flavors and the Iwasaki gauge action \\
}

\author{
T.~Kaneko$^{\, 1,2}$,
S.~Aoki$^{\, 3,4}$,
M.~Della Morte$^{\, 5}$, 
S.~Hashimoto$^{\, 1,2}$, 
\\
R.~Hoffmann$^{\, 6}$ and
R.~Sommer$^{\, 7}$ 
\\[5mm]
(CP-PACS/JLQCD and ALPHA Collaborations) \\[5mm]
\small 
$^1$
High Energy Accelerator Research Organization (KEK),
Tsukuba, Ibaraki 305-0801, Japan 
\\
\small
$^2$
Graduate University for Advanced Studies,
Tsukuba, Ibaraki 305-0801, Japan
\\
\small 
$^3$
Graduate School of Pure and Applied Sciences, \\
\small University of Tsukuba, Tsukuba, 
Ibaraki 305-8571, Japan 
\\
\small
$^4$
Riken BNL Reserach Center, Brookhaven National Laboratory, 
Upton, NY 11973, USA
\\
\small 
$^5$
CERN, Physics Department, TH Unit, 
CH-1211 Geneva 23, Switzerland
\\
\small
$^6$
Department of Physics, University of Colorado,
Boulder, CO 80309, USA
\\
\small 
$^7$
DESY, 
Platanenallee 6, 15738 Zeuthen, Germany
}

\date{\small 9 March 2007}

\begin{document}
\maketitle

\begin{abstract}

We perform a non-perturbative determination of the improvement 
coefficient $\cA$ to remove $O(a)$ discretization errors 
in the axial vector current in three-flavor lattice QCD with 
the Iwasaki gauge action and the standard 
O$(a)$-improved Wilson quark action.
An improvement condition with 
a good sensitivity to $\cA$ is imposed at constant physics. 
Combining our 
results with the perturbative expansion, $\cA$ is
now known rather precisely for $a^{-1}\!\gtrsim\!1.6$~GeV.
\end{abstract}


\maketitle

\newpage


\section{Introduction}

Various discretizations of QCD on a lattice are presently used 
in the large scale efforts aiming at non-perturbative results
in the theory of strong interactions 
(see \cite{lat06:Giusti,new:cern,Spectrum:Nf3:CPPACS-JLQCD:lat06,PACS-CS,lat06:KS,lat06:KS2,lat06:DWF,lat06:tm,new:tm,new:GW} and
references therein). 
Wilson's original formulation~\cite{Wilson} 
is theoretically very well founded~\cite{Reisz,luescher:tm} and rather
simple to implement in numerical simulations.
The flavor symmetries are exact and with modern 
algorithms~\cite{Hasenbusch-trick,PHMC:JLQCD,algo:L2,Urbach:2005ji,RHMC}
the regime of small quark masses and small lattice spacings can be reached
\cite{lat06:Giusti,algo:stability}.  
On the other hand it is well known that since 
the chiral symmetries are broken by the Wilson term,
lattice artifacts linear in the lattice
spacing are present. It has long been understood how these can be removed 
by applying Symanzik's improvement 
programme~\cite{Oa-improvement,SW_term,impr:theory:ALPHA,Oa-improvement:review1,impr:theory:nondeg}. 
One has to add 
dimension five fields to the lattice Lagrangian and (for example)
dimension four fields to the quark bilinears. In particular the bare flavor axial current
\be
   A_\mu^a(x) = \overline{\psi}(x)\, T^a \, \gamma_\mu\gamma_5\,\psi(x)
\ee  
(the SU($\nf$) generator $T^a$ acts in flavor space)
is improved by
\bea
   (A_{\rm I})_\mu^a(x) &=&  A_\mu^a(x) +
   a\,\ca\,\frac12(\partial_\mu+\partial_\mu^*)\,P^a(x)\,, \quad
   P^a(x) =\overline{\psi}(x)\, T^a \, \gamma_5\,\psi(x) \,, 
   \\[-1ex] \nonumber \text{with}\qquad\qquad&& \\[-1ex]
   \partial_\mu\,f(x) &=& \frac{1}{a} [f(x+a\hat{\mu})-f(x)]\,,\quad
    \partial^*_\mu\,f(x) = \frac{1}{a} [f(x)-f(x-a\hat{\mu})]\,.
\eea
The coefficients of these correction terms, such as $\ca$, can be determined 
non-perturbatively  by requiring specific
continuum chiral Ward-Takahashi identities to be valid at 
finite lattice spacing~\cite{NP_csw:Nf0:Plq:ALPHA}. One is then sure 
that the O$(a)$ effects are entirely removed. Details of this programme
as well as the present status have recently been 
reviewed~\cite{Oa-improvement:review2}. Here we just mention that 
the coefficient $c_{\rm sw}$ of the Sheikholeslami-Wohlert term \cite{SW_term},
the only dimension five correction to the action\footnote{We neglect small
O$(am)$ modifications of the gauge couplings and quark masses 
\cite{impr:theory:ALPHA,impr:theory:nondeg}, as they are not so relevant
in practice \cite{Oa-improvement:review2}. Here $m$ stands for any of the quark
masses.},
has been determined non-perturbatively for different gauge actions and 
numbers of
flavors~\cite{NP_csw:Nf0:Plq:ALPHA,NP_csw:Nf2:Plq:ALPHA,NP_csw:Nf2-3:Plq:CPPACS-JLQCD,NP_csw:Nf0-3:RG:CPPACS-JLQCD}.

Next to $c_{\rm sw}$, the axial current improvement coefficient $\ca$ is
of particular relevance -- for example in the determination of 
weak leptonic decay constants such as $F_\pi$ or the quark masses.  
Non-perturbative determinations
of $\ca$ have been studied for $\nf=0$ and $2$  
in refs.\cite{NP_csw:Nf0:Plq:ALPHA,NP_cA:Plq:BGLS:WI,NP_cA:Plq:UKQCD:Gap,NP_cA:Nf0:Plq:ALPHA:WF,NP_cA:Nf2:Plq:ALPHA:WF}.
It turned out that they need special care since the spread 
between $\ca$-values computed from different improvement conditions 
is significant around $a\approx 0.1$~fm. There is nothing fundamentally
wrong with this fact. However, 
as explained in some detail in 
refs.~\cite{Oa-improvement:review2,NP_csw:Nf2-3:Plq:CPPACS-JLQCD,NP_cA:Nf2:Plq:ALPHA:WF,impr:babp}, 
in such a situation 
it is important to impose improvement conditions on a line of constant
physics. This means that as the lattice spacing $a$ is varied, all other
physical scales are kept fixed. The remaining effects (after
improvement) are then smooth O($a^2$) terms.

Here we apply this strategy to the theory with $\nf=3$ flavors and
the Iwasaki gauge action \cite{Iwasaki_gauge}, which is of immediate 
interest to the large scale computations of 
the CP-PACS/JLQCD Collaborations~\cite{Spectrum:Nf3:CPPACS-JLQCD:lat06}.
All known practical methods for a computation of $\ca$ start from the fact
that in the continuum limit 
the (PCAC) quark mass 
         \be
            m =                   
            \frac{ \langle \phi' | \frac{1}{2}\left(\partial_\mu+\partial_\mu^{*}\right)
                            (A_I)_0^{a} | \phi \rangle }
                 { 2 \langle \phi' | P^{a} | \phi \rangle }
            \label{eqn:PCACmass}
         \ee
does not depend on the choice of the external
states $|\phi \rangle\,,\; |\phi' \rangle$. This is just a rephrasing of the
PCAC (operator) identity. On the lattice an O($a$) dependence will exist in
general. It is reduced to O($a^2$) by improvement. Requiring $m$ to be the 
same for two different choices of $|\phi \rangle\,,\; |\phi' \rangle$, or
as we will say later ``two different kinematical conditions'', allows 
a determination of  $\ca$ when $c_{\rm sw}$ is already known. 

As in ref.~\cite{NP_cA:Nf2:Plq:ALPHA:WF}, we use the Schr\"odinger functional
defined in a Euclidean $L^3\times T$ world 
to construct suitable states with a large sensitivity to $\ca$. 
In the following section we define the exact choices of kinematical
conditions. The reader who is familiar with ref.~\cite{NP_cA:Nf2:Plq:ALPHA:WF}
may skip this section and proceed directly to the description of the
simulation details, sect.~\ref{s:sim}, and the results, sect.~\ref{sec:results}. 
We finish with some conclusions.


\section{Improvement condition}
\label{sec:impr_cnd}

We introduce the following  Schr\"odinger functional~\cite{SF1,SF2} correlation functions
\cite{NP_cA:Nf0:Plq:ALPHA:WF,NP_cA:Nf2:Plq:ALPHA:WF}
\bea
            f_{\rm A}^{(n)}(x_0) 
            & = &
            -\frac{a^3}{3} \sum_{\bf x} \langle A_0^a(x) \calO^{a,(n)} \rangle\;,
            \quad
            f_{\rm P}^{(n)}(x_0) 
             = 
            -\frac{a^3}{3} \sum_{\bf x} \langle P^a(x) \calO^{a,(n)} \rangle
            \quad {\rm and}
            \label{fAP}
            \\
            f_1^{(n,m)}
            & = &
            -\frac{1}{3}\langle \calO^{\prime \, a, (n)} \calO^{a, (m)}\rangle,
            \label{f1}
         \eea
with
         \be
            \calO^{a,(n)}
             = 
            \frac{a^6}{L^3}\,
            \sum_{\bf y, z} \omega^{(n)}({\bf y}-{\bf z}) \, 
            \bar{\zeta}({\bf y}) \, T^a \, \gamma_5 \, \zeta({\bf z}).
            \label{enq:bound_op}
         \ee
where $\zeta$ and $\bar\zeta$ are the fermionic boundary fields on the $x_0=0$ timeslice ($\cal{O}'$
is defined in the same way in terms of the boundary fields at $x_0=T$). 
The correlators depend
on the smooth functions $\omega^{(n)}$.
Here, as in  
ref.~\cite{NP_cA:Nf2:Plq:ALPHA:WF}, we use three wave functions
\bea
   \omega^{(n)} 
   & = &
   \frac{1}{N^{(n)}} 
   \sum_{{\bf k} \in {\bf Z}^3}
   \bar{\omega}^{(n)}(|{\bf r}-{\bf k}L|)
   \hspace{5mm}
   (n\!=\!1,2,3),
   \\[2mm]
   \bar{\omega}^{(1)}({\bf r})  
   & \propto & 
   e^{-|{\bf r}|/a_0}\; , \quad
   \bar{\omega}^{(2)}({\bf r})  \propto  (|{\bf r}|/r_0)\, e^{-|{\bf r}|/a_0}\; , \quad 
            \bar{\omega}^{(3)}({\bf r})  \propto  e^{-|{\bf r}|/(2a_0)}\; ,
            \label{eqn:wave_func}
         \eea
         with $a_0\!=\!L/6$. 
The normalization factors $N^{(n)}$ are fixed by
$a^3 \sum_{\bf x} (\omega^{(n)})^2 \!=\! 1$.

By suitably combining the operators ${\cal{O}}^{a,(n)}$, the resulting correlation functions
get contributions from different states in the pseudoscalar channel.
In fact we construct the boundary operators $\calO_0$ and $\calO_1$, 
which mainly couple to the ground and first excited states respectively, by using the
eigenvectors of the $3 \times 3$ symmetric matrix $f_1^{(n,m)}$
         \be
            \calO_0^a
             = \sum_n \eta_0^{(n)} \calO^{a,(n)}, \quad
            \calO_1^a = 
            \sum_n \eta_1^{(n)} \calO^{a,(n)},
         \ee
where $\eta_0$ ($\eta_1$) represents the eigenvector associated
with the largest (2nd largest) eigenvalue. The corresponding correlators 
$f_{{\rm X},i}   =  \sum_n \eta_i^{(n)} f_{\rm X}^{(n)}$ with $i=0,1$ and
X$=$A,P are eventually used to define the improvement condition, which reads
         \be
            m_0(x_0) = m_1(x_0),
            \label{eqn:impr_cond}
         \ee
         where
         \bea
            m_i(x_0)        & = &             r_i(x_0) + c_{\rm A} \,a\, s_i(x_0), 
            \label{eqn:2state:mq}
	    \\
            r_i(x_0)             & = & 
            \frac {\left(\partial_0+\partial_0^*\right)\, f_{{\rm A},i}(x_0)}
            {4 f_{{\rm P},i}(x_0)} \quad {\rm and} \quad
            s_i(x_0) 
            = \frac{\partial_0 \partial_0^* \, f_{{\rm P},i}(x_0)}
                   {2\,f_{{\rm P},i}(x_0)      }.
         \eea
Solving eq.~(\ref{eqn:impr_cond}) for $c_{\rm A}$ yields
         \be
	    c_{\rm A} (x_0)
             = 
            - \frac{1}{a}\frac{\Delta r(x_0)}{\Delta s(x_0)},
            \hspace{5mm}
            \Delta r (x_0) = r_1(x_0)-r_0(x_0),
            \hspace{5mm}
            \Delta s (x_0) = s_1(x_0)-s_0(x_0)
            \;.\label{eqn:cA_eff}
	 \ee
The sensitivity of the improvement condition to $c_{\rm A}$ is given by  
$a\,|\Delta s(x_0)|$.
In the ideal case of exact projection on the ground ($\pi$) and first excited
($\pi^1$) states (and large $T-x_0$)
that would be given by $a\,(m_{\pi^1}^2-m_{\pi}^2)$. As discussed in 
ref.~\cite{NP_cA:Nf2:Plq:ALPHA:WF} however the vectors $\eta_i^{(n)}$ do not achieve perfect
projection  and  the correlator $f_{{\rm A},1}(x_0)$
for example gets some contribution from the ground state. 
Anyway what is really needed is that at intermediate times $x_0\simeq T/2$ where we extract 
$c_{\rm A}$, the correlation functions 
are dominated by  states with different energies, 
such that the sensitivity is high. 
We will see in sect.~\ref{sec:results} 
that in our setup this is indeed the case.


\section{Simulation details}
\label{s:sim}
We work in the theory with three (dynamical) degenerate flavors of  
non-perturbatively improved Wilson-fermions and
the Iwasaki gauge action~\cite{SW_term,NP_csw:Nf0-3:RG:CPPACS-JLQCD,Iwasaki_gauge}. 
The latter reads
         \bea
            S_g 
            & = &
            \beta
            \left\{\;\sum_{x,\,\mu <\nu} w_{\mu\nu}^P(x_0) 
                     \frac{1}{3}\, {\rm Re}\,\Tr [ 1\!-\!P_{\mu\nu}(x) ]
                    +\sum_{x,\,\mu ,\,\nu} w_{\mu\nu}^R(x_0) 
                     \frac{1}{3}\, {\rm Re}\,\Tr [ 1\!-\!R_{\mu\nu}(x) ] \right\},
            \label{eqn:Iwasaki}
         \eea
         where $\beta\!=\!6/g_0^2$, and $P_{\mu\nu}$ and $R_{\mu\nu}$ are 
         the $1 \times 1$ and $1 \times 2$ Wilson loops in the $(\mu,\nu)$ 
         plane.
         Their weights are 
         $w_{\mu\nu}^P\!=\!3.648$ and $w_{\mu\nu}^R\!=\!-0.331$ on periodic
         lattices.


    The Schr\"odinger functional formalism is implemented on a $L^3 \times T$ lattice with $L\!=\!T$. 
The background field is set to zero and the fields are chosen to be periodic
in space.
The weights in the gauge action are modified to the following choice~\cite{PTcA1}
         \bea
            w_{\mu\nu}^P(x_0) 
            & = & 
            \left\{
            \begin{array}{ll} 
               1/2 &  
               \mbox{at $x_0\!=\!0$ or $T$, and $\mu,\nu\!\ne\!4$} \\
               3.648   &
               \mbox{otherwise}
            \end{array}
            \right.
            \\
            w_{\mu\nu}^R(x_0) 
            & = & 
            \left\{
            \begin{array}{ll} 
               0                 &  
               \mbox{at $x_0\!=\!0$ or $T$, and $\mu,\nu\!\ne\!4$} \\
               -0.331\times(3/2) &  
               \mbox{at $x_0\!=\!0$ or $T$, and $\mu\!=\!4$} \\
               -0.331            &
               \mbox{otherwise}
            \end{array}
            \right.
          \eea
which entails tree level O($a$) improvement ``at the boundaries''~\cite{impr:theory:ALPHA}.
The coefficients 
of the O($a$) boundary counterterms for the fermions are also set to their
tree level value~\cite{impr:theory:ALPHA}. Note that this is not at all
essential. Irrespective of whether the boundary improvement terms are 
implemented, eq.~(\ref{eqn:impr_cond}) is a correct improvement condition~\cite{impr:theory:ALPHA}.

We simulate at three points in the $(\beta,L/a,\kappa)$ space on a line of constant physics 
defined by keeping the volume and the quark mass fixed. 
Scales are fixed through
$r_0$~\cite{r0} and we will use $r_0=0.5$~fm to quote physical units. 
For our action, the ratio $r_0/a$ has been computed 
in the region $1.83 \leq \beta \leq 2.05$ \cite{Spectrum:Nf3:CPPACS-JLQCD:lat06}\footnote{In
ref.~\cite{Spectrum:Nf3:CPPACS-JLQCD:lat06} $r_0/a$ is extrapolated to the
physical point, with the strange quark mass determined from the physical mass
of the Kaon.}. With a slight interpolation of the data of 
ref.~\cite{Spectrum:Nf3:CPPACS-JLQCD:lat06} we fixed $L/r_0 \approx 3$,
somewhat larger than the 
physical size used in ref.~\cite{NP_cA:Nf2:Plq:ALPHA:WF}.
The resulting
pairs $(\beta,L/a)$, together with some algorithmic details are collected in 
table~\ref{tbl:simparam}.

\begin{table}[bhtp]
\begin{center}
\begin{tabular}{ccccccccc}
   \hline
   $\beta$ & $L/a$ & $\kappa$ & $N_{\rm MD}$  & $N_{\rm poly}$ & $N_{\rm traj}$ 
                         & $P_{\rm PHMC}$ & $P_{\rm NMT}$  \\
   \hline
   1.83 & 12 & 0.13852 &  90 & 200 & 3800 & 0.90 & 0.97 \\
   1.83 & 12 & 0.13867 &  90 & 220 & 3800 & 0.89 & 0.95 \\ 
   1.95 & 16 & 0.13685 & 125 & 230 & 3000 & 0.91 & 0.96 \\
   1.95 & 16 & 0.13697 & 140 & 260 & 3000 & 0.94 & 0.94 \\
   2.05 & 20 & 0.13604 & 130 & 350 & 3000 & 0.87 & 0.97 \\ 
   \hline
\end{tabular}       
\caption{\small \it
   Simulation parameters. We denote the number of the Molecular Dynamics 
   steps by $N_{\rm MD}$, the order of the polynomial approximation 
   by $N_{\rm poly}$ and the number of unit length trajectories
   by $N_{\rm traj}$. 
   The acceptance rates for the PHMC updating 
   and for the noisy Metropolis test in the PHMC algorithm are denoted
   by $P_{\rm PHMC}$ and $P_{\rm NMT}$ respectively.
}
\label{tbl:simparam}
\end{center}
\end{table}
The hopping parameter $\kappa$ is tuned in order to give a bare quark mass
$m_{\rm ref}$ of about 15~MeV. At $\beta\!=\!1.83$ and $1.95$,
two quark masses around 10 and 20~MeV are simulated so that we can 
interpolate $\cA$ to $m_{\rm ref}$.
Notice that we are ignoring presumably small changes of the renormalization 
factors in our range of $\beta$ and we just keep the bare quark mass fixed.
Also, the 1-loop value of $\cA$~\cite{PTcA1,PTcA2} is used at this point in the 
definition of the quark mass.

The algorithm has been described in ref.~\cite{PHMC:JLQCD}. It is a
combination of HMC~\cite{HMC} and PHMC~\cite{PHMC1,PHMC2}. Non-Hermitian
Chebyshev polynomials $P[D]$ are used to approximate the inverse square root
of the Dirac operator $D$ required for the third flavor, whereas
the other two flavors are treated using the usual HMC pseudo-fermion
action. 
The number of molecular dynamics steps is chosen such that 
the acceptance rate $P_{\rm PHMC}$ is about 90\%.
In order to make this algorithm exact, the correction factor
         \bea
            P_{\rm corr} = \det\left[ W[D] \right],
            \hspace{5mm}
            W[D] = P[D]D
         \eea
is taken into account by a noisy Metropolis test~\cite{NMT}.
The order of the polynomial approximation is chosen 
such that the acceptance rate of the noisy Metropolis test is around 95\%.
Throughout all the computation the symmetrically even-odd preconditioned
version of the Dirac operator~\cite{PHMC:JLQCD,SHMC} is used.

The correlators in eqs.~(\ref{fAP}, \ref{f1}) are measured each 5th trajectory and
residual autocorrelations are estimated by binning the jack-knife samples.
For $c_{\rm A}(x_0)$ the errors flatten out for bin-sizes larger than four,
which is what we finally use in our analysis.


\section{Numerical results}
\label{sec:results}

\subsection{Wave function projection}

\begin{table}[!b]
\begin{center}
\begin{tabular}{cc|cc}
   \hline
   $\beta$ & $\kappa$   & $\eta_0$ & $\eta_1$ \\
   \hline
   1.83  & 0.13867 & ( 0.5459,  0.5920,  0.5929 ) 
                   & ( 0.8323,  -0.3019, -0.4649 ) \\
   1.95  & 0.13697 & ( 0.5415,  0.5942,  0.5948 )
                   & ( 0.8367,  -0.312\phantom{0},   -0.4500  ) \\
   2.05  & 0.13604 & ( 0.5360,  0.5962,  0.5976 )
                   & ( 0.8371, -0.2836,  -0.4679  ) \\
   \hline
\end{tabular}
\caption{\small \it
   Example of eigenvectors $\eta_0$ and $\eta_1$ in three-flavor QCD
   at each $\beta$.
}
\label{tbl:evec}
\end{center}
\end{table}

As discussed above, the analysis starts with the determination of the
eigenvectors of the correlator matrix $f_1^{(n,m)} (n,m=1,2,3)$. The results
at the lightest quark mass for each $\beta$ value are given in table~\ref{tbl:evec}.
The errors on the components of the eigenvectors are less than $10^{-3}$ and
$10^{-2}$ for $\eta_0$ and $\eta_1$, respectively.

The entries of the (normalized) eigenvectors are ratios
of correlation functions for which the $Z$-factors of the boundary fields
cancel. They will thus have a well defined continuum limit along
a line of constant physics as long as the wave functions $\omega^{(n)}$ are defined only 
through physical length scales.
Indeed, we observe only a small lattice spacing dependence and also note that the
values we obtain are close to those in ref.\cite{NP_cA:Nf2:Plq:ALPHA:WF}, where slightly
smaller values of $a_0$ ($\sim\!0.2$~fm) and $L$ are used in two-flavor QCD with the plaquette
gauge action.

\begin{figure}[htbp]
\vspace*{10mm}
\begin{center}
	 \includegraphics[width=76mm,clip]{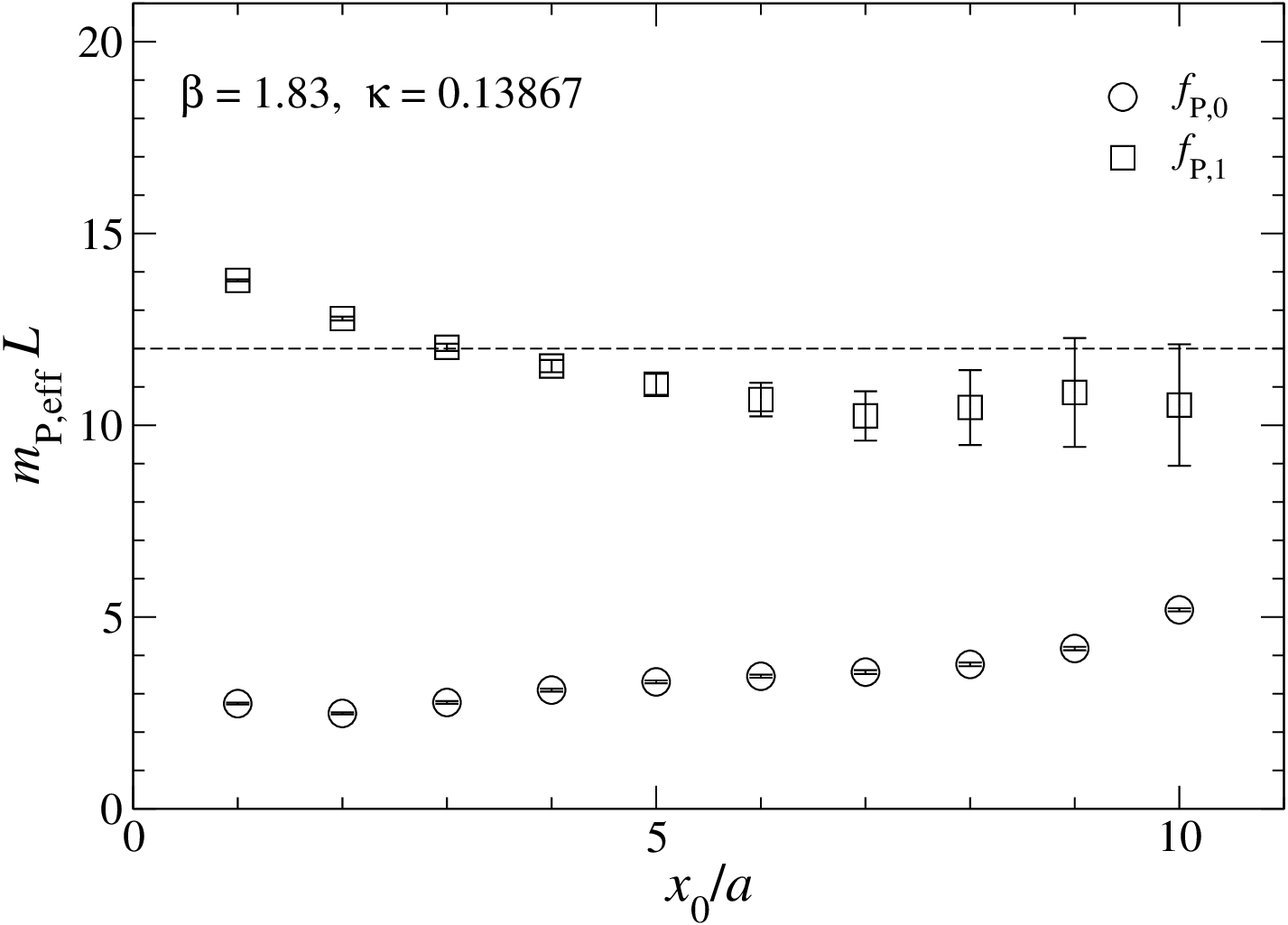}
   \hspace*{2mm}
   \includegraphics[width=77mm,clip]{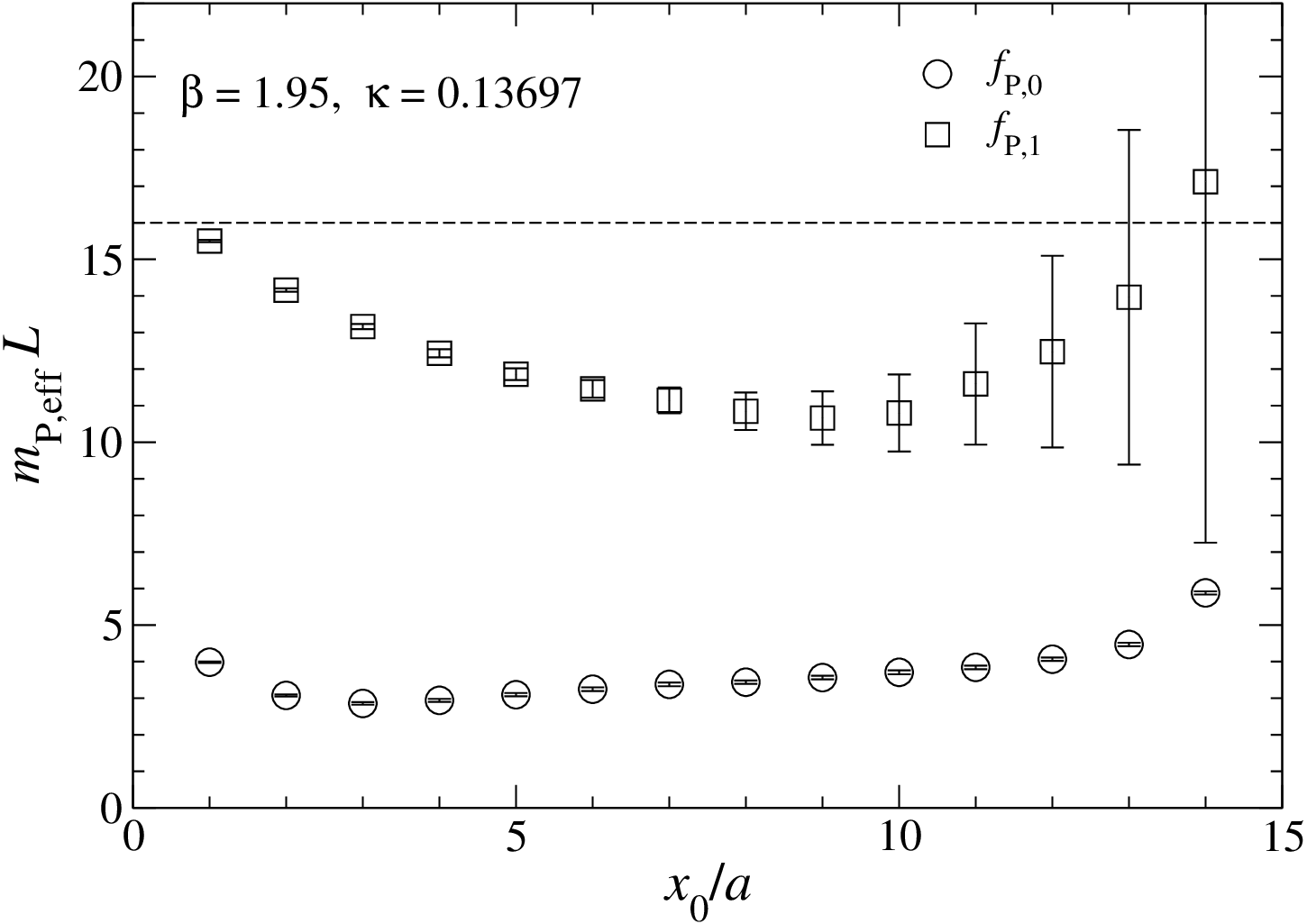}   
   \caption{\small \it
      Effective mass 
      $m_{\rm P,eff} = \frac{1}{2a} \log(f_{{\rm P},i}(x_0\!-\!a)/f_{{\rm P},i}(x_0\!+\!a))$ 
      for $f_{\rm P,0}$ and $f_{\rm P,1}$
      rescaled by $L$.
      The dotted line shows where the effective mass is equal to $a^{-1}$.
   }
   \label{fig:Nf3:emPL}
\end{center}
\end{figure}

Using the measured eigenvectors, we now construct the pseudoscalar correlators
$f_{{\rm X},0}$ and $f_{{\rm X},1} \ (\rm X\!=\!P,A)$. Fig.~\ref{fig:Nf3:emPL} shows the effective
masses in units of the box size for the projected correlators $f_{{\rm P},i}$ at
$\beta=1.83$ and 1.95. Clearly, the correlators are dominated by different states
and the effective masses are well separated even for large times. The data at the two
coarser lattice spacings are obtained at physical
quark masses similar to each other and we observe good agreement also
for the effective masses in the pseudoscalar channel
($m_{\rm P,0}\,L\!\approx\!3$ and $m_{\rm P,1}\,L\!\approx\!11$). In
table~\ref{tbl:drs} we have included also the combination 
$L^2\,\Delta s$, which has a continuum limit when all
quantities are computed on a line of constant physics. While
an $a$-dependence appears to be present in this combination, 
this is small. We can take its smallness 
as good evidence that our improvement condition
does not suffer from large O$(a^2)$ contributions.

Note, however, that the effective mass
$m_{{\rm P},1}$ at our smallest $\beta$ is already close to the cutoff
(i.e. $m_{{\rm P},1}\,L \sim L/a$ in fig.~\ref{fig:Nf3:emPL}).
This implies that a rapid increase of the residual ${\rm O}(a^2)$
effects might occur if one were to evaluate the improvement condition
at even coarser lattice spacings. 
    
\subsection{The improvement coefficient}

With the projected correlation functions we can proceed to the extraction
of $\cA$ itself. Table~\ref{tbl:drs} lists the differences $\Delta r$
and $\Delta s$ for the lightest quark mass at each $\beta$. 
In all cases we see a good signal for $\Delta s$ and thus have a large 
sensitivity to $\cA$.

\begin{table}[htbp]
\begin{center}
\begin{tabular}{cc|ccc}
   \hline
   $\beta$ & $\kappa$     & $a\,\Delta r$ & $a^2\,\Delta s$ & $L^2\,\Delta s$ \\
   \hline
   1.83  & 0.13867   & 0.0229(14)            & 0.429(22)           & 62(3)\\
   1.95  & 0.13697   & 0.0072(\phantom{1}7)  & 0.236(14)           & 60(4)\\
   2.05  & 0.13604   & 0.0036(\phantom{1}3)  & 0.133(\phantom{1}6) & 53(2)\\
   \hline
\end{tabular}
\caption{\small \it
   Examples of $\Delta r$ and $\Delta s$ at $x_0\!=\!T/2$.
}
\label{tbl:drs}
\end{center}
\end{table}

\begin{figure}[htbp]
\vspace*{10mm}
\begin{center}
   \includegraphics[width=73.5mm,clip]{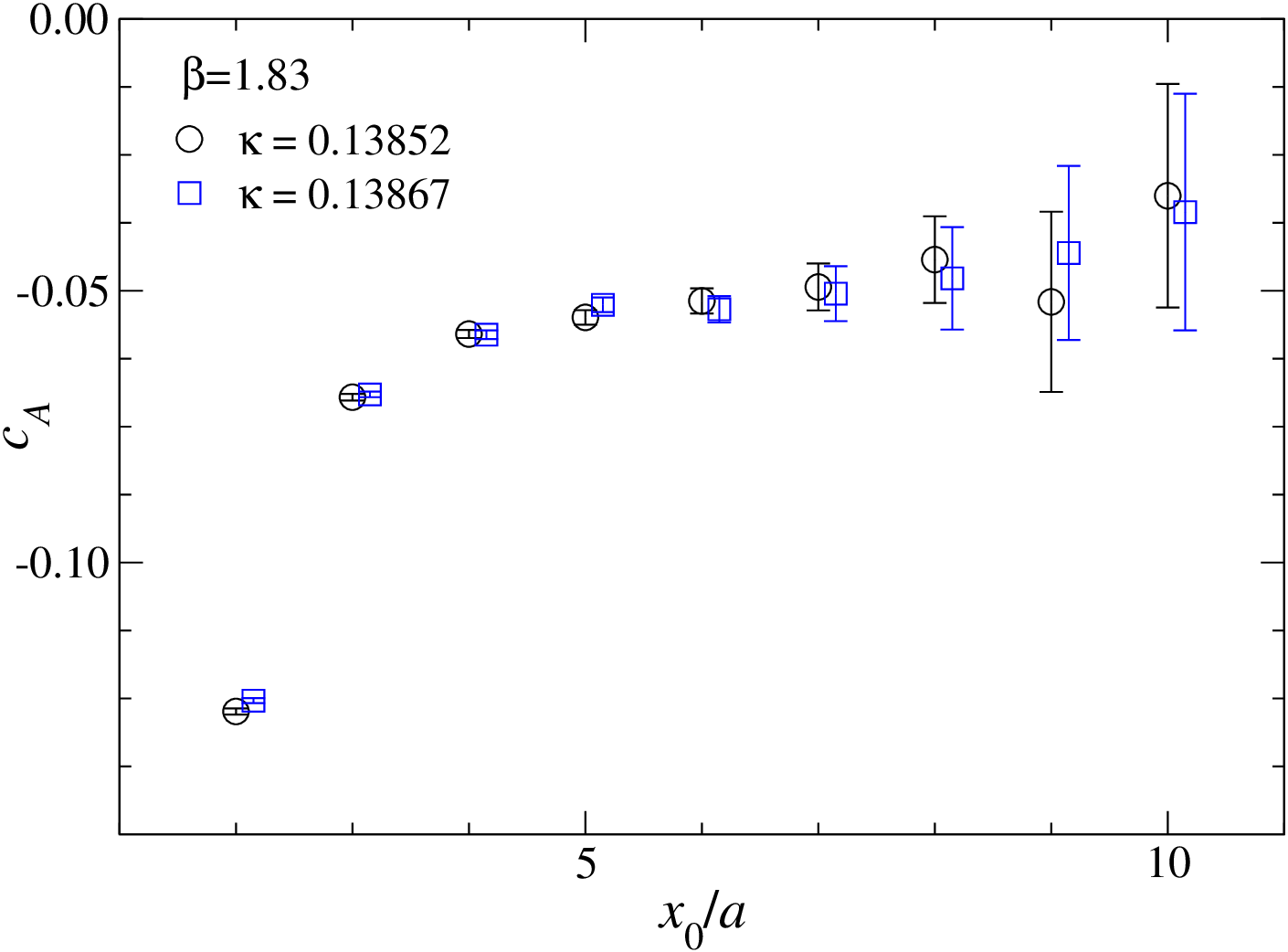}
   \hspace{2mm}
   \includegraphics[width=76mm,clip]{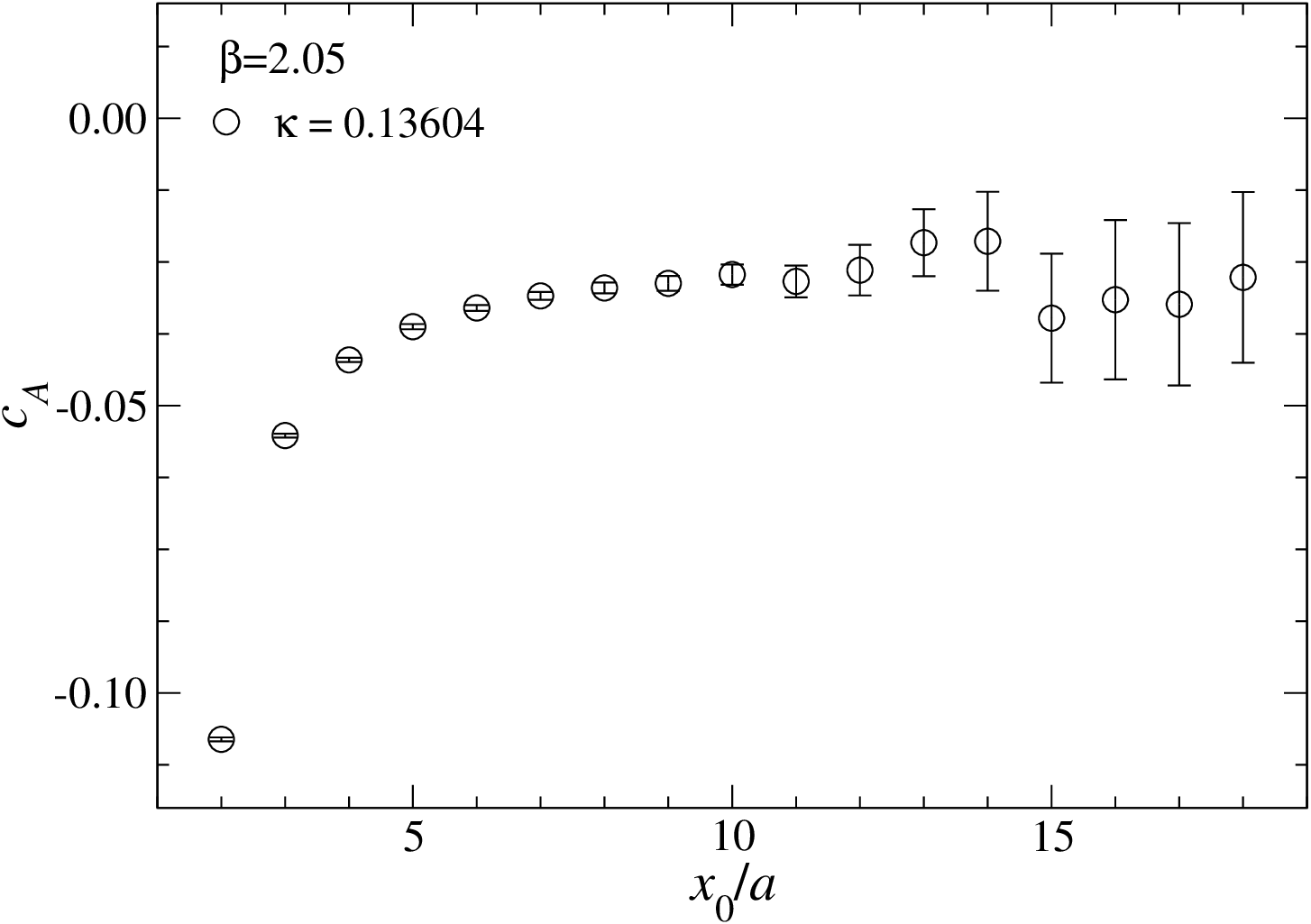}
   \caption{\small \it
      Effective value of $\cA$ as a function of $x_0$ 
      at $\beta\!=\!1.83$ (left panel) and $\beta\!=\!2.05$ (right panel).
      The data at $\beta\!=\!1.83$ and $\kappa\!=\!0.13867$ are slightly shifted
      along the vertical axis for clarity.
   }
   \label{fig:Nf3:cA}
\end{center}
\end{figure}

In fig.~\ref{fig:Nf3:cA}, we plot the effective $\cA(x_0)$, cf. eq. (\ref{eqn:cA_eff}),
from the finest and coarsest lattices. The value of $\cA(x_0)$ stabilizes
after only a few lattice spacings from the lower temporal boundary
where higher exited states contribute. 
In all cases $x_0=T/2$
is already in the region, where these effects are small and we use this choice
to complete the definition of $\cA$. Results for the improvement coefficient and the
PCAC quark mass from all simulations are collected in table~\ref{tbl:cA:sim_mq}.
         
\begin{table}[htbp]
\begin{center}
\begin{tabular}{lc|cl}
   \hline
   $\ \ \beta$ & $\kappa$     & $am$       & $\quad \ \cA$       \\ 
   \hline
   2.05    & 0.13604 &  0.00554(14) & -0.0272(18) \\
   \hline
   1.95    & 0.13685 &  0.01020(29) & -0.0348(25) \\
       		 & {\small [interp.]}   & $am_{\rm ref}$   & -0.0319(18)\\
          & 0.13697 &  0.00508(28) & -0.0303(24) \\   
   \hline
   1.83    & 0.13852 &  0.01406(54) & -0.0519(23) \\
   & {\small [interp.]}   & $am_{\rm ref }$   & -0.0528(17)\\
           & 0.13867 &  0.00614(63) & -0.0534(24) \\   
   \hline
\end{tabular}
\caption{ \small \it
   Numerical results for $am$ and $\cA$. Also shown are the results of the
   interpolation to $m_{\rm ref}$ at the two coarser lattice spacings.
}
\label{tbl:cA:sim_mq}
\end{center}
\end{table}

\subsection{Interpolation of $\cA$}

As discussed above, we aim at evaluating the improvement condition on a line
of constant physics in order to avoid potentially large O($a$) ambiguities
in $\cA$ itself. To this end we interpolate the results for $\cA$ at
$\beta=1.83$ and 1.95 to a quark mass $m_{\rm ref}$ that is matched to the one
measured on the finest lattice. The quark mass dependence seems to be very small
and thus the uncertainties in the quark masses become unimportant and we obtain
$\cA$ at $m_{\rm ref}$ with a small statistical error.

\begin{figure}[htbp]
\begin{center}
   \includegraphics[width=80mm,clip]{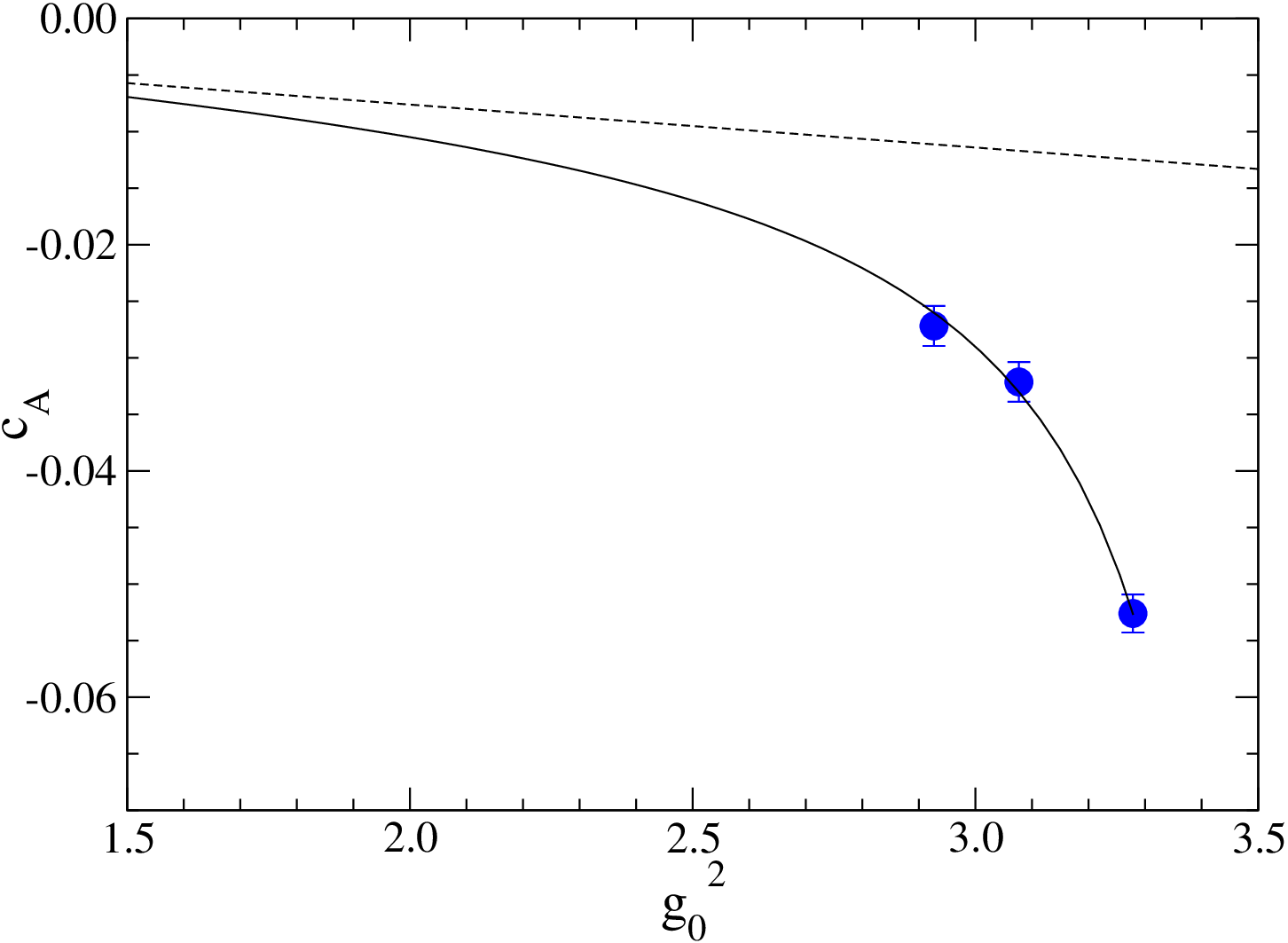}
   \caption{\small \it
      Non-perturbative estimate of $\cA$ as a function of $g_0^2$.
   }
   \label{fig:cA_vs_g2}
\end{center}
\end{figure}

For future use we summarize the present results for the improvement coefficient
in an interpolating formula (\ref{eqn:cA_vs_g2}), which by construction
reduces to the one-loop result from refs.~\cite{PTcA1,PTcA2} in the perturbative limit
\bea
  \cA(g_0^2) = -0.0038\, g_0^2 \, \times \, \frac{1 - 0.195\,g_0^2}{1 - 0.279\,g_0^2}\;.
  \label{eqn:cA_vs_g2}
\eea
It is plotted in fig~\ref{fig:cA_vs_g2}, where one can verify that this formula reproduces
the data well and gives a smooth interpolation in the range of $\beta$ values we simulated.
As was the case with the plaquette gauge action and two quark flavors
\cite{NP_cA:Nf2:Plq:ALPHA:WF}, the non--perturbative result is quite different
from the one-loop estimate for practically relevant lattice spacings.

\vspace{5mm}

\subsection{Systematic uncertainties}

The computation of $\cA$ on a line of constant physics reduces the intrinsic 
ambiguity on the improvement coefficient to a smooth
O($a$) form. Deviations from this condition will lead
to systematic effects and we should therefore check the consequences of variations
of the physical volume and quark mass for our improvement condition.

All simulations, on which we report here, are performed at fixed physical
volume and we thus have no direct check of the volume effects on $\cA$
from this improvement condition. 
From \cite{NP_cA:Nf2:Plq:ALPHA:WF} we know that those can be large, but
we know that our condition guarantees
that they disappear smoothly as we approach the continuum limit, especially
since our volume scaling is
based on actual measurements of $r_0/a$.

From the data at the two coarser lattice spacings in table~\ref{tbl:cA:sim_mq}
it is evident that the quark mass dependence of $\cA$ is very weak in our setup.
This implies that no fine tuning of $m$ is required and also a posteriori
justifies the fact that we ignore small 
changes
of the renormalization factor in our range of $\beta$
and use the bare quark mass in our definition of a line of
constant physics.

As mentioned above, the energy of the first excited state
at our lowest $\beta\!=\!1.83$ is close to $a^{-1}$. Consequently,
enforcing the present improvement condition at $\beta\! \lesssim \!1.83$
may induce large $O(a^2)$ scaling violations in the axial current.
While larger volumes might help in lowering this energy, the observation shows
that even with improved gauge actions,
one should not push the simulations too much towards coarse lattice 
spacings. 
On the other hand it is useful to repeat our earlier observation:
within the range of lattice spacings covered here, we see 
a reasonable scaling of $L^2\,\Delta s$; this is a good hint
that the considered matrix elements do {\em not} suffer from large 
$a$-effects. In retrospect the same statement can be made about the
$\nf=2$ computation with plaquette gauge action \cite{NP_cA:Nf2:Plq:ALPHA:WF}.


\section{Conclusions}

We have computed the O($a$)-improvement coefficient  $\cA(g_0)$  of the axial current 
non-per\-tur\-ba\-tively in three-flavor QCD with the Iwasaki gauge 
action and non-perturbative $c_{\rm sw}(g_0)$ \cite{NP_csw:Nf0-3:RG:CPPACS-JLQCD}.
The improvement coefficient $\cA$ is parametrized as a function of $g_0$. 
Since the results 
connect smoothly to the one-loop formula at weak 
coupling, a simple interpolation formula (\ref{eqn:cA_vs_g2})
could be given in the range of 
$a^{-1} > 1.6$~GeV. 

We note that at the largest lattice spacing 
covered, the correction term amounts to 10\,--\,15\% in decay constants
$F_\pi,\;F_K$ and then also in the renormalized
quark masses evaluated from the PCAC relation. 
As a next step, a full non-perturbative evaluation of these 
quantities now requires the computation of the renormalization
factor $Z_{\rm A}$ and for interesting applications of vector form factors
also the corresponding quantities $Z_{\rm V}$ and $c_{\rm V}$ are very 
relevant.
On the other hand, improvement terms proportional to the
light quark masses are suppressed by the smallness of $am$. It then 
appears sufficient to approximate the associated coefficient by
one-loop perturbation theory~\cite{PTcA1,PTcA2}.

\section*{Acknowledgment}

We thank Stephan D\"urr for useful discussions.
Numerical simulations are performed on Hitachi SR8000
at High Energy Accelerator Research Organization (KEK)
under a support of its Large Scale Simulation Program (No.~05-132).
This work is supported by the Grant-in-Aid of the
Ministry of Education, Culture, Sports, Science and Technology
of Japan (No.~13135204, 15540251, 17740171, 18034011, 18340075) 
and the JSPS Core-to-Core Program.
TK is grateful to the Theory Group in DESY Zeuthen 
for kind hospitality during his stay.


\end{document}